\documentstyle[graphics,epsf]{lamuphys}
\makeatletter
\let\chapter\hid@chapter
\makeatother

\def \etal {{\rm et al.}}
\def \eg {{\em e.g.}}

\def\eps@scaling{.95}
\def\epsscale#1{\gdef\eps@scaling{#1}}

\def\plotone#1{\centering \leavevmode
    \epsfxsize=\eps@scaling\columnwidth \epsfbox{#1}}

\begin{document}
\pagenumbering{arabic}

\title{The Deepest Spectrum of the Universe?
  Constraints on the Lyman Continuum Background at High Redshift}
\titlerunning{The Deepest Spectrum of the Universe?}
\author{Andrew~J.~Bunker\inst{1,2}, Francine~R.~Marleau\inst{1,2} and
  James~R.~Graham\inst{2}}
\institute{
Institute of
Astronomy, Madingley Road, Cambridge CB3 0HA, UK\\
{\tt email: bunker@ast.cam.ac.uk}
\and
Department of
Astronomy, University of California at Berkeley,\\ 601 Campbell Hall,
Berkeley CA~94720, USA}

\maketitle

\begin{abstract}
We describe an ongoing experiment to search for the meta-galactic
Lyman-continuum background at $z\approx 2 - 3$.  We are obtaining one of
the deepest optical spectra ever, using LRIS/Keck-II to search for the
fluorescent Ly-$\alpha$ emission from optically thick H{\scriptsize~I}
clouds.  The null results of our pilot study (Bunker, Marleau \& Graham
1998) placed a $3\sigma$ upper bound on the mean intensity of the
ionizing background of $J_{\nu0} < 2\times
10^{-21}$\,erg\,s$^{-1}$\,cm$^{-2}$\,Hz$^{-1}$\,sr$^{-1}$ at $z \approx
3$.  This constraint was more than two orders of magnitude more
stringent than any previously published direct limit. Our results
excluded the possibility that decaying relic neutrinos are responsible
for the meta-galactic radiation field.  We have recently greatly
extended our search, obtaining a 16-hour spectrum which is sensitive to
UV background fluxes $\approx
10^{-21}$\,erg\,s$^{-1}$\,cm$^{-2}$\,Hz$^{-1}$\,sr$^{-1}$ ($z\approx
2.3$ at 3\,$\sigma$, assuming the H{\scriptsize~I} clouds are $\approx
10$\arcsec\ in extent).  We describe how the results of this study can
be used to constrain the quasar luminosity function and the contribution
of high-redshift star-forming galaxies to the ambient ionizing
background.
\end{abstract}


\section{Introduction}

The meta-galactic ultraviolet (UV) background has a central
astrophysical r\^{o}le.  The absence of a Gunn-Peterson trough in the
spectra of $z \approx 5$ QSOs implies that the inter-galactic medium
(IGM) must have been highly ionized by the UV background at even greater
redshift.  This Lyman continuum background is responsible for
maintaining the Ly$\alpha$ forest clouds in a highly-ionized state, and
may also cause the sharp edges of H{\scriptsize~I} disks in nearby
spirals (Dove \& Shull 1994).  It is widely believed that the
meta-galactic UV flux is the integrated light of QSOs, or hot massive
stars in young galaxies, or both.  However, this ionizing background has
never been directly detected.  We present here an overview of an ongoing
observational program designed to achieve this goal.

\section{Seeking the Ultraviolet Background}

Hogan \& Weymann (1987) proposed that long-slit spectroscopy of ``blank
sky'' should reveal patches of fluorescent Ly$\alpha$ emission, excited
by the meta-galactic ionizing background, from the population of clouds
whose absorption produces the Ly$\alpha$ forest in QSO spectra.  A
measurement of the surface brightness of this fluorescent emission puts
limits on the incident ionizing flux at high redshift.  Gould \&
Weinberg (1996) present a treatment of the transport of Ly$\alpha$ in
clouds with $N_{\rm H\,I} = 10^{17} - 10^{20}$\,cm$^{-2}$.  For
optically thick clouds ($\tau_{912} > 5$, $N_{\rm H\,I} >
10^{18}$\,cm$^{-2}$) the flux of Ly$\alpha$ photons from recombination
cascades is equal to 0.6 times the flux of incident ionizing photons;
this fraction is robust and independent of cloud geometry. The
Ly$\alpha$ photons are absorbed and re-emitted until scattering from an
atom with a velocity $(v - \bar{v}) \approx \pm 4 \sigma$, at which
point it can escape. A typical Ly$\alpha$ cloud with a velocity
dispersion of $\sigma = 30$\,km\,s$^{-1}$ (Kim \etal\ 1997) would have
double-peaked fluorescent Ly$\alpha$ line with a width of
240\,km\,s$^{-1}$.  Hence, moderate resolution spectroscopy
($\lambda/\Delta \lambda_{\rm FWHM} \ga 1000$) of an optically thick
cloud, known to exist from a QSO absorption system, gives a direct
measurement of the energy in the ionizing background. Moreover, since
there are a typically one or more Lyman-limit systems per sight-line at
$z\sim 2-4$, the same long-slit exposure would also detect tens of
serendipitous clouds making it possible to make a two-dimensional (2D)
map of the Ly$\alpha$ forest.

\section{Our Survey}

The calculations of Gould \& Weinberg (1996) suggest that it should be
possible to detect fluorescent Ly$\alpha$ emission from optically thick
Ly$\alpha$ clouds at $z \sim 3$ with a deep ($>$~10~hour) long-slit
spectrogram on a 10-m telescope.  Motivated by this, we have embarked on
an extremely sensitive spectroscopic search with the Keck~II Telescope.
Our pilot study (detailed in Bunker, Marleau \& Graham 1998) showed that
it is possible to reach the required line fluxes with the Low Resolution
Imaging Spectrograph (LRIS, Oke \etal\ 1995).  Over the past year we
have greatly extended our program, obtaining a total of 16~hours of
integration time.  This constitutes one of the deepest optical spectra
ever obtained.  

The data for our extended program was taken towards the quasar\\
DMS~2139.0-0405 (Hall \etal\ 1996) and a 1\arcsec-wide long-slit was
used with a blue-blazed grating of resolving power $\lambda/\Delta
\lambda_{\rm FWHM} \approx 2000$.  Our observations sample the
wavelength range 3750--5500\,\AA, corresponding to Ly$\alpha$ in the
redshift range $2.1 < z < 3.5$. Our most recent 8~hours of data were
obtained using 4 parallel slits with a 10\% bandwidth filter
($\lambda_{\rm cent}=4000$\,\AA ) to cover a much larger solid angle
while concentrating on those redshifts ($z \approx 2.3$) where our
sensitivity to $J_{\nu 0}$ is greatest (see Fig.~2).

\section{Data Analysis}

The details of the data reduction are presented in Bunker, Marleau, \&
Graham (1998).  Great care was taken in removing the spectrum of the
night sky while preserving any cosmological signal (expected to be
spatially-extended line emission of low surface brightness, with a
velocity spread $< 500\,{\rm km\,s}^{-1}$). Rather than doing the
sky-subtraction in the usual manner by fitting a polynomial to each
column (which might subtract the extended emission we are looking for),
we first rectified the sky lines using a distortion matrix.  We then
subtracted off a high signal-to-noise ratio (SNR) sky spectrum from each
detector row, scaled to the slit illumination at that point along the
slit.  We search for extended Ly$\alpha$ emission by smoothing along the
spatial axis of the background-subtracted composite 2D spectrum (\eg,
Figs.~1b\,\&\,1c) on various scale lengths between the size of the seeing disk
($\la 1\arcsec$) and the length of the slit (3\arcmin --7\arcmin).  To
amplify any signal present in our data, we calculate the 2D power
spectrum. This may potentially reveal the combined signal for a
population of clouds that are too faint to be detected individually (see
the simulation in Fig.~1d). Finally, we use the SExtractor algorithm
(Bertin \& Arnouts 1996) to catalogue objects and we determine the
completeness of our search method via artificial-clouds experiments.

\begin{figure}[h]
\resizebox{\textwidth}{!}{\hspace{1.2cm}(a)\hspace{2cm}(b)\hspace{2cm}(c)\hspace{2cm}(d)\hspace{1.2cm}}
\plotone{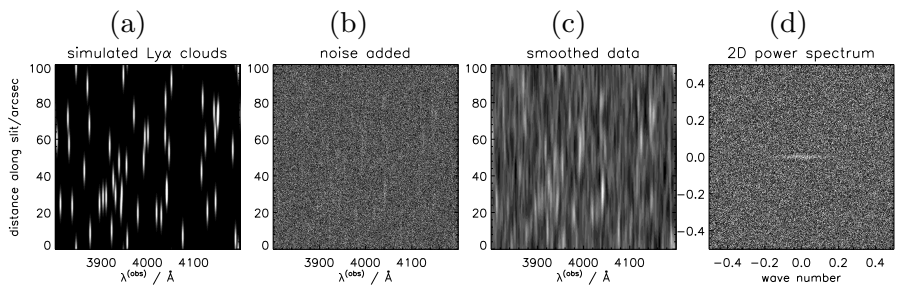}
\caption{A simulation of an LRIS long-slit 2D spectrum containing 50
clouds, each with a FWHM in the dispersion direction of
$250$\,km\,s$^{-1}$ and a length of 10\arcsec\ (Fig.~1a).  Noise is
added to the image with the peak of a cloud corresponding to ${\rm SNR}
= 1$ per resolution element (Fig.~1b), which renders individual clouds
invisible to casual inspection by eye.  However, when the noisy image is
convolved with a 2D Gaussian kernel of comparable size to the clouds,
their signal is revealed (Fig.~1c).  The 2D power spectrum of the noisy
image is calculated in order to combine the signal of the population of
clouds (the central peak in Fig.~1d).  Simulations show that when single
clouds are not visible, the combination of many clouds is enough to make
the detection possible.}
\end{figure}

\section{Our Limits on the UV Background}

Our initial study (Bunker, Marleau \& Graham 1998) based on a 1.5\,hour
LRIS Keck spectrum failed to find Ly$\alpha$ fluorescence from
optically-thick clouds.  A lack of Ly$\alpha$ emission constrains the UV
flux.  The upper limit on the ambient UV background at $2.7 < z < 3.1$
(where our constraints on $J_{\nu 0}$ from the pilot study are most
stringent) is equivalent to a flux at the Lyman limit of $J_{\nu 0} <
2.0 \times 10^{-21}$\,erg\,s$^{-1}$\,cm$^{-2}$\,Hz$^{-1}$\,sr$^{-1}$
(Fig.~2).  This assumes optically-thick clouds with dimensions $\approx
10\arcsec$ -- the size of such systems derived from Mg{\scriptsize~II}
QSO absorber studies (Steidel \& Dickinson 1995).  This limit on $J_{\nu
0}$ is almost two orders of magnitude lower than any previous direct
limit (Lowenthal \etal\ 1990; Mart\'{\i}nez-Gonz\'{a}lez \etal\ 1995)
but is still three times above the expected contribution of known QSOs
for $q_{0}=0.5$ (Haardt \& Madau 1996).  This implies that the
completeness of optical QSO catalogs is better than 30\% and that the
contribution to $J_{\nu 0}$ at $z \approx 3$ from star-forming galaxies
(Songaila \etal\ 1990) cannot exceed twice that from known QSOs.  We
calculate that the escape fraction of Lyman continuum photons from
star-forming galaxies at these redshifts must therefore be less than
10--50\% (depending on the dust obscuration of the rest-UV continuum for
$z\approx 3$ galaxies).  Based on our extended program, our sensitivity
thus far on the ambient UV background at $z \sim 2.3$ probes down to a
flux at the Lyman limit of $J_{\nu 0} \la
10^{-21}$\,erg\,s$^{-1}$\,cm$^{-2}$\,Hz$^{-1}$\,sr$^{-1}$ (see Fig.~2).

\begin{figure}[h]
\plotone{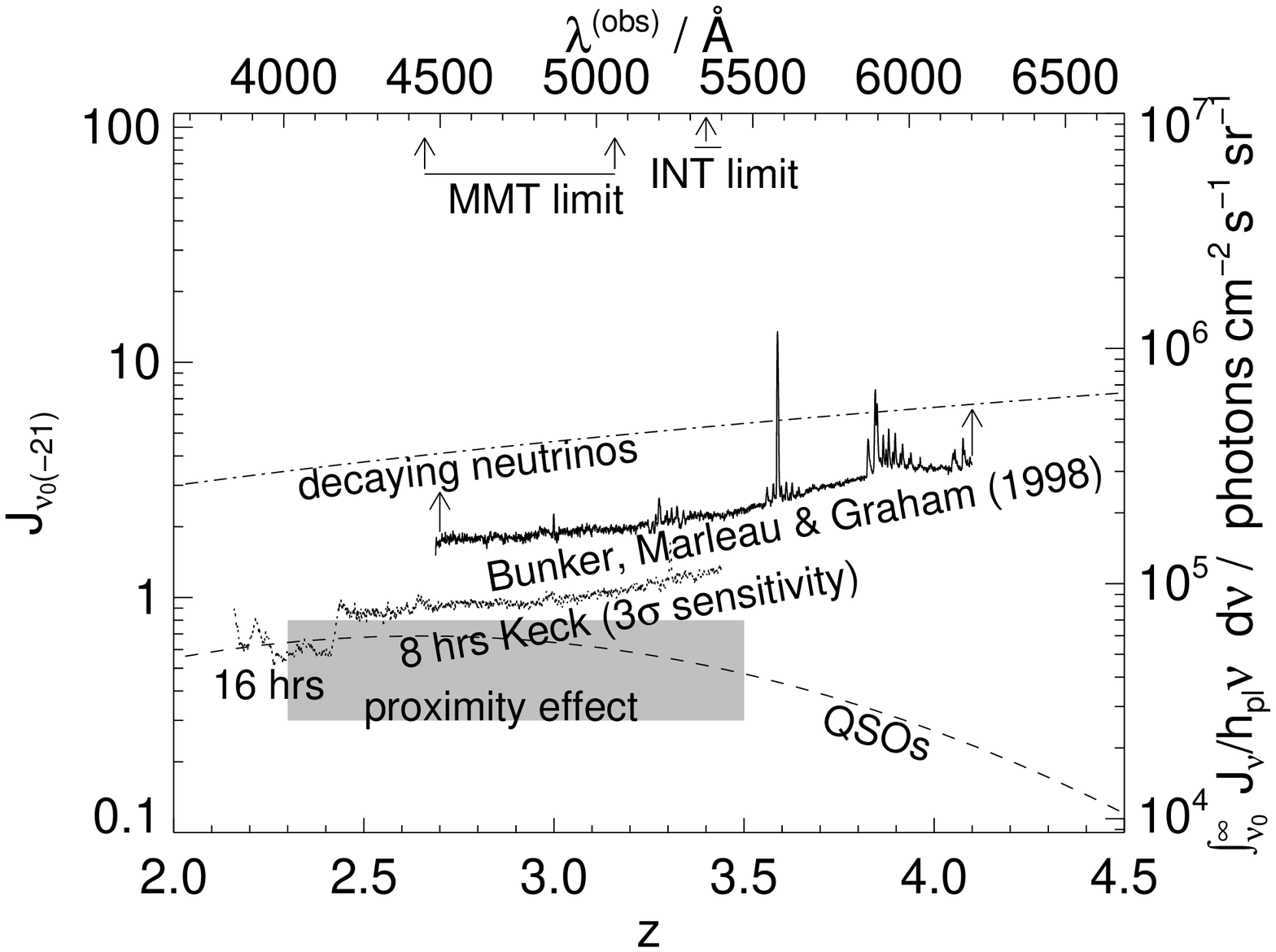}
\caption{Our 3\,$\sigma$ upper-limits on the flux density at the Lyman
edge,
$J_{\nu0(-21)}=J_{\nu0}~/~10^{-21}$~erg~s$^{-1}$~cm$^{-2}$~Hz$^{-1}$~sr$^{-1}$,
as a function of redshift (solid line) from our 1.5-hour pilot study
(Bunker, Marleau \& Graham 1998).  The null results of this preliminary
search mean that the region above the heavy solid curve is excluded.
Also shown are the previous 3\,$\sigma$ upper-limits from the INT
(Mart\'{\i}nez-Gonz\'{a}lez 1995) and the MMT (Lowenthal \etal\ 1990),
converted to the same assumed projected cloud size ($\approx
10\arcsec$).  We also plot the 3\,$\sigma$ sensitivity reached in our
recently extended program: we have obtained a 16~hour spectrum with
LRIS/Keck-II (dotted curved), and we are currently processing this
enlarged data set to look for the signature of the ionizing background,
reprocessed as Ly$\alpha$.  The shaded region is the lower-limit on
$J_{\nu0}$ from the QSO proximity effect (Espey 1993). The estimated
contribution to $J_{\nu0}$ from the luminosity function of known
high-$z$ QSOs is also shown as the dashed line (Haardt \& Madau
1996). The equivalent Lyman continuum flux from decaying relic neutrinos
is plotted as a dot-dash line, adopting the parameters of Sciama (1998)
and assuming these neutrinos form the bulk of the dark matter in an
$\Omega_{\rm M}=1$ Universe. Such a scenario is strongly ruled out by
the null results of our pilot study, as we would have detected
Ly$\alpha$ fluorescence from the flux of ionizing decay photons.}
\end{figure}

\section{Conclusion}

A search with LRIS/Keck-II is used to constrain the fluorescent
Ly$\alpha$ emission at $z \approx 2 - 3$ from the clouds which produce
the higher-column-density component of the Ly$\alpha$ forest.  The null
results of a pilot study by Bunker, Marleau \& Graham (1998) provided
the best upper limit yet on the ionizing UV background of $J_{\nu 0}\la
2\times 10^{-21}$\,erg\,s$^{-1}$\,cm$^{-2}$\,Hz$^{-1}$\,sr$^{-1}$
(3\,$\sigma$ limit at $2.7 < z < 3.1$).  We have now extended our
integration time to 15~hours, attaining a 3\,$\sigma$ sensitivity of
$J_{\nu 0} \la 10^{-21}$\,erg\,s$^{-1}$\,cm$^{-2}$\,Hz$^{-1}$\,sr$^{-1}$
at $z\approx 2.3$ -- one of the deepest optical spectra ever
obtained. We are currently conducting a power-spectrum analysis to
detect the signature of the ionizing background at high redshift.

\section*{Acknowledgments}

We thank the Max Planck Institute for Astronomy, particularly Hans
Hippelein and Klaus Meisenheimber, for an enjoyable and informative
meeting at Ringberg. We are grateful to everyone at the Keck observatory
for their assistance with our program. AJB and FRM acknowledge support
from the Cambridge Institute of Astronomy PPARC observational rolling
grant, ref.~no.~PPA/G/O/1997/00793. AJB acknowledges a NICMOS
postdoctoral fellowship while at Berkeley (grant NAG\,5-3043). JRG is
supported by the Packard Foundation.

\end{document}